\documentclass{article}
\usepackage{graphicx}
\begin{document}
\title{Spontaneous Wireless Networking to Counter Pervasive Monitoring}
\author{E. Baccelli, O. Hahm, M. W\"ahlisch \footnote{Emmanuel Baccelli is affiliated with INRIA and Freie Universit\"at Berlin. Oliver Hahm is affiliated with INRIA. Matthias W\"ahlisch is affiliated with Freie Universit\"at Berlin.}}
\maketitle

\section{Pervasive Monitoring is a Myth}
Pervasive monitoring does not exist for one good reason: if monitoring does happen, as revealed by recent whistleblowing that highlighted the scale of NSA surveillance activities ~\cite{Snowden}, it happens in very few key central locations in the network. Monitoring, as we know it so far, is thus by no means pervasive in the geographical or topological sense. One might further speculate that such monitoring will probably never become pervasive because it would simply be too expensive to put in place. \ \\ \ \\
This observation leads to a fundamental question: how could NSA surveillance reach such an incredibly large scope, to the point where it \emph{seems} pervasive, while in fact, they only monitor the network from a few select locations? The answer to this question is simple: the NSA's monitoring could scale because \emph{we} are addicted to centralization. \ \\ \ \\
From the physical point of view, we made ourselves absolutely dependent on the deployed, fixed infrastructure. Whenever we communicate, the quasi totality of the traffic relies upon going through a central access point of some sort. This approach to networking certainly offers substantial advantages (a good performance/cost compromise for starters), but on the other hand it offers cheap and efficient vantage points for whoever seeks large scale monitoring opportunities. It does not have to be so.\ \\ \ \\
Similarly, from the logical point of view, we make ourselves more and more dependent on cloud services. Whenever we store, or access our data, we rely on a centralized architecture that concentrates everything in data centers. Again, this approach offers great advantages (e.g. easier access from anywhere, lower cost reliability and efficiency), but on the other hand it offers more cheap and efficient vantage points for would-be spies. It does not have to be so either.
\section{Making Massive Surveillance Harder}
At this point, it becomes apparent that there are essentially three ways to make large scale systematic surveillance a much harder task. \ \\ \ \\
The first type of solution is to significantly increase the physical security of potential vantage points. While this type of solution is obviously necessary, history shows that there are always cracks to infiltrate if the will to break in is there. And it will be there. So this type of solution is by no means sufficient.\ \\ \ \\
The second type of solution is to significantly increase the default levels of encryption used over the network. For starters, the default behavior should be to encrypt "door-to-door" and don't let  communication be decrypted by intermediate devices. Furthermore the default cryptographic mechanisms should be significantly harder to decipher. Note that to ensure this actually happens consistently, the use of open source software should be mandatory concerning such cryptographic mechanisms. On one hand, as highlighted in ~\cite{blog}, open-source bundled with automatic updates can slow down or stop obsolescence altogether, and on the other hand, open source is best known guarantee against malware and potential backdoors. However, if one of the physical end points of the communication is a vantage point of potentially high interest (say a data center), we quickly come back to the problems of the first type of solution, i.e. there are always cracks to infiltrate if the will to break in is there. And it will be there, with an extremely intense focus, because this is a potential vantage point -- and with an intense focus one can do wonders.\ \\ \ \\
The third type of solution is of an entirely different nature, which is the main purpose of this document. Instead of hardening the current architecture and increasing the security of high profile targets, another category of approach could be to aim for \emph{target dispersal}, as suggested recently by B. Schneier in ~\cite{Schneier}. Target dispersal would eliminate "default" vantage points and thus naturally disable systematic mass surveillance. De facto, surveillance efforts would be forced to be more specific and personalized, and thus more directly accountable for. \ \\ \ \\
To that end, \emph{stateless} networking approaches should be developed and employed, i.e. approaches that do not rely on central entities (infrastructure-based), but rather on spontaneous interaction between autonomous peer entities, as locally as possible (topologically and/or geographically). The network architecture should evolve towards a mode of operation where all possible stateless solutions are tried first, before considering any infrastructure-based approach -- again, the goal being to increase target dispersal. In fact, several examples of stateless approaches have appeared over the last decade, at different layers. These include for instance peer-to-peer (P2P) networks or WebRTC \cite{Web-RTC} at the application layer, and multi-hop spontaneous wireless networks at the network layer ~\cite{Sigcomm-Chapter}, such as mobile ad hoc networks \cite{RFC2501}, wireless sensor networks, vehicular networks, or wireless mesh networks.\ \\ \ \\
While P2P networks have been massively deployed and adopted (WebRTC will likely enjoy the same fate) spontaneous wireless networks have not yet been widely adopted. There has however been quite some work in the research community over the past decade on this topic, which has identified issues that must (still) be tackled in order to efficiently adapt IP protocols for these new networks. Actually, the community has already produced a number of new protocol specifications dedicated to spontaneous wireless networking operation, such as \cite{RFC3626} \cite{RFC3561} \cite{RFC5449}, \cite{RFC6550} among others. Some of these protocols have even been deployed in specific contexts. Several european and american military applications have deployed using mobile ad hoc routing to power spontaneous, on-the-move local communications between elements on-site. In another context, several cities in Europe and in the USA have deployed wireless community mesh networks (e.g. \cite{FREIFUNK} \cite{FUNKFEUER}), which use these protocols. Last, but not least, the highly anticipated Internet of Things (IoT) is heavily based on spontaneous wireless networking and the outcome of efforts such as 6LoWPAN \cite{6LOWPAN-WG} or ROLL \cite{ROLL-WG}.\ \\ \ \\
Indeed, merging spontaneous wireless networks with traditional, operated networks presents a number of advantages, aside of target dispersal to counter massive surveillance. First, it could offer operator infrastructure offloading \cite{MANIAC13}. Second, the resulting network would systematically maximize connectivity at marginal cost. Third, it offers natural coverage extension for the network access infrastructure already deployed. And last, it offers increased resilience of the network in face of infrastructure outage.\ \\ \ \\
There are of course lingering issues that do not allow off-the-shelf, spontaneous wireless networking at large scale. Among others, such issues include (i) the absence of efficient standard router auto-configuration schemes in this context \cite{RFC5889}, which has profound implications, (ii) the lack of dedicated, optimized link layer technology (which has focused on infrastructure based networking for decades), (iii) no standard key exchange protocol or alternatives for efficient authentication method to date, and (iv) some fundamental differences in terms of characteristics  compared to traditional networks, such as throughput capacity \emph{vs} number of nodes in the network.\ \\ \ \\
However, in light of the recent events revealing massive and systematic surveillance \cite{Snowden}, there is a strong argument to decentralize our networking paradigms, and in this realm, spontaneous wireless networking at the networks layer is an asset that should not be overlooked. Contrary to infrastructure-based approaches, which are prone to monitoring, spontaneous wireless networking uses communication links that are local and volatile, i.e. unless one is physically present at the time and location of the communication, one must abandon all hope of monitoring anything. While the wireless nature of the communication may on the other hand facilitate eavesdropping, the fact that one has to be there at the time/place of the communication significantly hampers mass surveillance, and cryptographic techniques can provide a privacy equivalent to what is achieved on wire. The honest, technical question that we should ask ourselves at this point is: are the lingering issues of spontaneous wireless networking integration in the IP stack worth solving/improving or not, taking into account our goal to maximize surveillance target dispersal?

\section{Position: More Spontaneity Please}
The position highlighted in this paper is that in order to maximize target dispersal to counter systematic massive surveillance, we should not overlook the full potential of decentralized network paradigms. This full potential includes not only mantras such as "\emph{think hard before you use the cloud}", or "\emph{let's bypass the service provider}" but also alternative techniques at the network layer, such as multi-hop spontaneous wireless networking, which could significantly reduce our dependence on the infrastructure -- which will remain, whether we want it or not, a desirable vantage point for mass monitoring. There are a number of issues that need to be resolved or alleviated towards native and massive integration of spontaneous wireless networks in the currently deployed IP architecture and infrastructure-based networks (massive IPv6 adoption could already help somewhat). However, the gains obtained in terms of target dispersal alone could be worth this cost -- and there are other gains too.

\end{document}